\documentclass[12pt,a4paper,final]{iopart}

\usepackage{iopams}  
\usepackage{graphicx}
\usepackage[breaklinks=true,colorlinks=true,linkcolor=blue,urlcolor=blue,citecolor=blue]{hyperref}
\begin{document}

\title[Various Topological Mott insulators in strongly-interacting 
boson system]{Various Topological Mott insulators and topological bulk charge pumping in strongly-interacting 
boson system in one-dimensional superlattice}

\author{Yoshihito Kuno$^{1}$, Keita Shimizu$^{2}$ and Ikuo Ichinose$^{2}$}
\address{$^1$ Department of Physics, Graduate School of Science, Kyoto University, Kyoto 606-8502, Japan}
\address{$^2$ Department of Applied Physics, Nagoya Institute of Technology, Nagoya 466-8555, Japan}
\ead{ykuno@yagura.scphys.kyoto-u.ac.jp}



\begin{abstract}
In this paper, we study a one-dimensional boson system in a superlattice potential. 
This system is experimentally feasible by using ultracold atomic gases, 
and attracts much attention these days. 
It is expected that the system has a topological phase called topological 
Mott insulator (TMI).
We show that in strongly-interacting cases,
the competition between the superlattice potential and 
the on-site interaction leads to various TMIs with non-vanishing 
integer Chern number. 
Compared to hard-core case, the soft-core boson system exhibits 
rich phase diagrams including various non-trivial TMIs. 
By using the exact diagonalization, 
we obtain detailed bulk-global phase diagrams including the TMIs with high 
Chern numbers and also various non-topological phases.
We also show that in adiabatic experimental setups, the strongly-interacting bosonic 
TMIs exhibit the topological particle transfer, i.e., topological charge pumping phenomenon,
similarly to weakly-interacting systems.
The various TMIs are characterized by topological charge pumping as 
it is closely related to the Chern number, and therefor the Chern number
is to be observed in feasible experiments. 
\end{abstract}

\vspace{2pc}
\maketitle

\section{Introduction}

Recent years, topological phase is one the most interesting 
subjects in condensed matter physics.
It is generally defined as a state characterized by a nontrivial topological
number even though it has no local order parameters.
Topological phase is expected to form even in one dimensional (1D) system
as suggested in the celebrated work by Thouless \cite{Thouless}. 
The topological phase in 1D system originates 
from geometrical similarity of the (1+1)D spacetime to 
2D space, in which well-known topological states of matter, e.g.,
the integer quantum Hall (IQH) \cite{TKNN, Kohmoto} 
and fractional quantum Hall (FQH) states \cite{Laughlin,Wen} form.
Inspired by the important observation by Thouless \cite{Thouless}, certain
1D models have been studied from the view point of topological 
phase \cite{Zhu,Deng, Lang, Matsuda,Xu,Ganeshan,Hu,Roscilde}.
Also the experiments on cold atomic gases in an optical lattice have started 
to ``quantum simulate" such 1D systems \cite{Lewenstein}. 
As one of the recent remarkable successes in the experiments, we notice
the realization of the topological Thouless pumping \cite{Nakajima, Lohse}. 
The topological Thouless pumping is a phenomenon in which 
a spontaneous atomic transportation takes place by changing
a Floquet parameter characterizing topological properties of the Hamiltonian.
The experimental successes stimulate theoretical study of the topological phase 
of 1D cold atomic system in an optical lattice. 

Motivated by these theoretical observations and experimental successes, 
we shall study the systems of {\em strongly-interacting Bose gases on 1D superlattice}
in this work.
In experiments, interactions between cold atoms can be controlled
by using optical experimental techniques, 
e.g., Feshbach resonance \cite{optical,Feshbach} and orbital Feshbach resonance \cite{OrbitalFeshbach}.
The target system is described by the Bose-Hubbard model (BHM) with an applied
modulate chemical potential term.
Interestingly, the BHM with a modular chemical potential 
is expected to have non-trivial topological states 
\cite{Zhu,Deng, Lang, Matsuda,Xu, Ganeshan} 
and it can be quantum simulated by the recent experiments \cite{Chen}.

Most of the previous studies have focused on the existence 
of non-trivial topological phases. 
Some numerical studies by using the density-matrix renormalization group method
(DMRG) confirmed the existence of the non-trivial topological phase
called topological Mott insulator (TMI) \cite{Zhu,Deng,Matsuda}. 
TMI is classified by a topological number such as the Chern number and it
has a gap in the bulk but a gapless excitation in the (spacial as well as 
phase diagram) boundaries.
A quantum Monte-Carlo simulation was also carried out to detect the topological phase \cite{Li}.
However, the existence was verified only in a limited parameter regime, and
detailed global phase diagrams are still lacking.
In particular, strongly-correlated bosonic topological states with {\it high Chern number} 
in the bulk have not been clarified yet in a global parameter regime, 
nor it is understood well how the competition between the superlattice
potential and the on-site repulsion determines the ground state of the system. 
Also, there is one important question, i.e., how the topological phases in the obtained
phase diagrams are related with the topological charge pumping as bulk topological
properties.

In this paper, we shall study the above problems in the strongly-interacting boson 
system by using the exact diagonalization \cite{ED1,ED3}, and 
show explicitly relation between the equilibrium topological phases 
and the topological charge pumping in the adiabatic process by following 
Ref.\cite{Hatsugai}.
From the relation, the global phase diagram plays a role of a guide for 
detecting various topological charge pumping in the experiments.

This paper is organized as follows. 
The target boson model in the 1D optical superlattice is explained in Sec.~\ref{Sec.II}.
Section~\ref{Sec.III} studies the ground states of the system for the hard-core 
boson limit. 
We explain the single particle (SP) equation, which is related
to the famous Harper equation in 2D electron lattice model in uniform 
magnetic fields, and we present the SP spectrum. 
We observe the ground states and their topological order 
by using the exact diagonalization. 
In Sec.~\ref{Sec.IV}, we clarify the ground state properties of the soft-core boson
system, where the on-site interaction plays an important role to determine 
the ground states and their topological properties. 
In particular, we shall show global phase diagrams of the Chern number. 
The phase diagrams includes rich topological phases. 
In Sec.~\ref{Sec.V}, we discuss a relationship between the ground states obtained 
by the exact diagonalization and the dynamical topological charge pumping. 
We shall show the Chern number 
of the many-body interacting ground-state is directly connected 
to the particle transfer performed in an adiabatic pumping cycle of superlattice.
Section~\ref{Sec.VI} is devoted for conclusion. 
Detailed calculations concerning to the Chern number and the topological charge pumping
are given in appendices.


\section{Bose-Hubbard model on 1D superlattice}\label{Sec.II}
\begin{figure}[t]
\centering
\includegraphics[width=6.5cm]{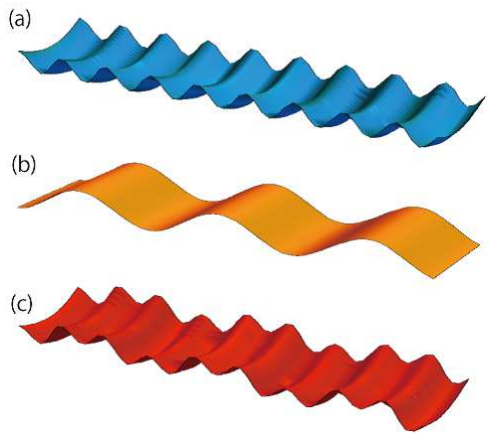}
\caption{Optical superlattice: 
(a) Standard optical one-dimensional periodic potential,
 which creates the tight bounding constraint for the system.
(b) Another lattice potential, described by adding
the periodic chemical potential term to the BHM.
(c) Superposed potential, called superlattice.
}
\label{Fig1}
\end{figure}

We consider a dilute Bose gas system in a superlattice,
whose Hamiltonian is given as follows,
\begin{eqnarray}
H_{\rm BH}&=&-\sum_{i}\biggl[Je^{i\theta /L}a^{\dagger}_{i}a_{i+1}+\mbox{h.c}\biggr]
+\frac{U}{2}\sum_{i}n_{i}(n_{i}-1)\nonumber\\
&&+V_{0}\sum_{i}\cos(2\pi \alpha i+\delta )n_{i},
\label{BHM1D}
\end{eqnarray}
where $a_{i} \ (a^\dagger_i)$ is the boson annihilation (creation) operator at 
superlattice sites $i$, the density operator $n_{i}=a^{\dagger}_{i}a_{i}$,
$J$ is the hopping amplitude, $U$ is the on-site repulsion.
The superlattice is created in cold atom experiments 
by using two different standing-wave lasers as shown in Fig.~\ref{Fig1}.
The parameter $V_{0}$ in $H_{\rm BH}$ [Eq.(\ref{BHM1D})]  is related to the amplitude of 
the superlattice potential.
The parameter $\alpha$ is the superlattice period, which is a tunable parameter in experiments, 
and we call $\alpha$ modulate parameter. 
On the other hand, $\delta$ is the phase shift between the two standing-wave lasers. 
$\theta$ is a twisted phase coming from the twisted boundary condition \cite{Rousseau,Niu} 
and $L$ is the system size.
In this paper, we consider the 1D system with the periodic boundary condition.
From the view point of the topological state,
adiabatic Floquet parameters of the present system are $\theta$ 
and $\delta$, i.e.,
the Hamiltonian $H_{\rm BH}$ [Eq.(\ref{BHM1D})] is invariant under 
the transformations of the adiabatic parameters such as 
$\theta\rightarrow \theta+2\pi$ and $\delta\rightarrow \delta+2\pi$,
independently.
From this fact, the adiabatic parameters span a 2D
periodic parameter space, i.e., a torus $T^{2}_{\theta\delta}$. 
In Sec.~5, we shall study the topological charge pumping, which takes place by varying
the parameter $\delta$ as a function of time $t\in [0,T]$ such as  $\delta(t)=2\pi t/T$.

\section{Phase diagrams of hard-core Bose-Hubbard model}\label{Sec.III}

In this section, we shall consider the hard-core boson limit $U \to \infty$, 
i.e., multiply occupied states are prohibited, and then we drop
the on-site interaction term in $H_{\rm BH}$. 
In this limit,
the system approaches to a non-interacting fermionic system \cite{Jordan}.
Under this condition, the SP spectrum of the model determines the
ground state of the system.
The previous studies \cite{Zhu,Deng,Lang,Xu} showed
that the SP spectrum of the present system
is given by the solutions of the Harper equation, 
i.e., the Hofstadter butterfly \cite{Hofstadter}, 
which describes the 2D lattice electron system in uniform magnetic fields.
Interestingly, this correspondence can be understood by the consideration of
the dimensional extension from spacial 1D system to spacial 2D system \cite{Kraus1}. 
Actually, by substituting the SP wave function 
$|\Psi\rangle_n = \sum_{i}\psi_{n,i}a^{\dagger}_i|0\rangle$ into 
$H_{\rm BH}|\Psi\rangle=E|\Psi\rangle$ with the BHM Hamiltonian of Eq.(\ref{BHM1D}),
the SP equation is obtained as follows,
\begin{eqnarray}
-J(\psi_{i+1,n}+\psi_{i-1,n})+V_{0}\cos(2\pi\alpha i+\delta)\psi_{i,n}
=E_{n}\psi_{i,n}.  \nonumber\\
\label{SingleE}
\end{eqnarray}
Then, let us compare the above SP equation (\ref{SingleE})
with the Harper equation \cite{Hofstadter}, 
$$
-t_{x}(\phi_{i+1,n}+\phi_{i-1,n})-2t_{y}\cos(2\pi f i-k_{y})\phi_{i,n}=E_{n}(k_{y})\phi_{i,n},
$$ 
where $t_{x(y)}$ is the hopping in the $x(y)-$direction and $f$ is the magnitude of 
the applied magnetic flux per plaquette.
As shown in Ref.~\cite{Kraus1},
the modulate parameter $\alpha$ and the phase shift $\delta$ in $H_{\rm BH}$ 
correspond to $f$ and the $y$-component wave number $k_{y}$, respectively. 
Furthermore, there exists correspondence such as
$t_{x}\longleftrightarrow J$,  $-2t_{y}\longleftrightarrow V_{0}$.
As a future work, it is interesting to study the above correspondence 
from a relativistic field-theoretical view point \cite{Kuno}.

\begin{figure}[t]
\centering
\includegraphics[width=9cm]{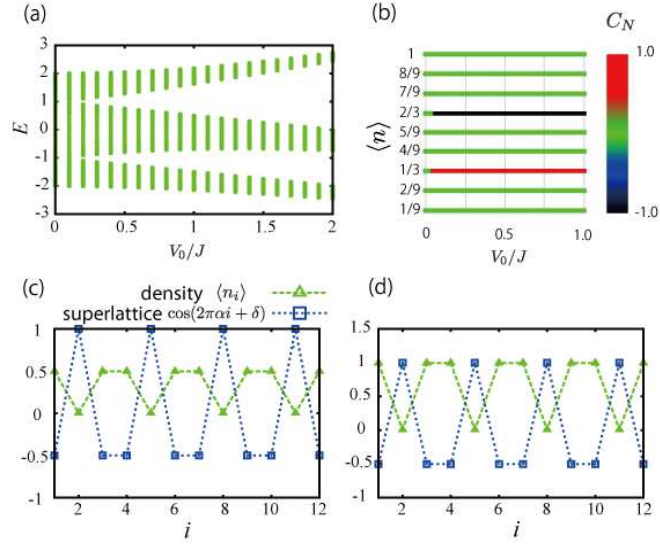}
\caption{(a) Energy spectrum obtained from 
the single particle equation for $\alpha=1/3$ and $\delta=5\pi/3$. 
As increasing the value of $V_{0}$, 
the energy levels split into the three bands for the case $\alpha=1/3$.
(b) $(\langle n\rangle $ - $V_{0}/J)$-plane phase diagram of the Chern number $C_{N}$.
Density snapshot for $V_{0}/J=1$, $\langle n\rangle=1/3$ (c) and $2/3$ (d) with $\delta=5\pi/3$. 
Each ground state is non-degenerate. 
From the density pattern, it is seen that the discrete translational invariance is broken.
The ground state has a 
nonzero-integer Chern number, $C_{N}=+1 \: (-1)$. $\alpha=1/3 \: (2/3)$.}
\label{Fig2}
\end{figure}

From the SP equation of the BHM of Eq.(\ref{SingleE}), it is naturally expected 
that there exist a band insulator at specific filling factors, 
[filling factor is an average particle number per site]. 
As an example, Fig.~\ref{Fig2} (a) shows the energy spectrum $E_n$ obtained
by solving Eq.(\ref{SingleE}) for the case of $\alpha=1/3$, $\delta=5\pi/3$
with various $V_0/J$.
We observe that as increasing $V_{0}/J$, the SP 
spectrum splits into three bands, i.e., the spectrum has two band gaps,
and each band has the $1/3$-filling.
Then in the hard-core boson system with $\langle n\rangle=1/3$, 
the first band is fully occupied, and a band insulating forms.
Similarly for the system with $\langle n\rangle=2/3$.
[See later discussion and Figs.~\ref{Fig2} (c) and (d).] 
It is known that the band insulating ground state has non-trivial topological nature
indicated by a non-vanishing integer Chern number \cite{TKNN}.

\begin{figure}[t]
\centering
\includegraphics[width=14cm]{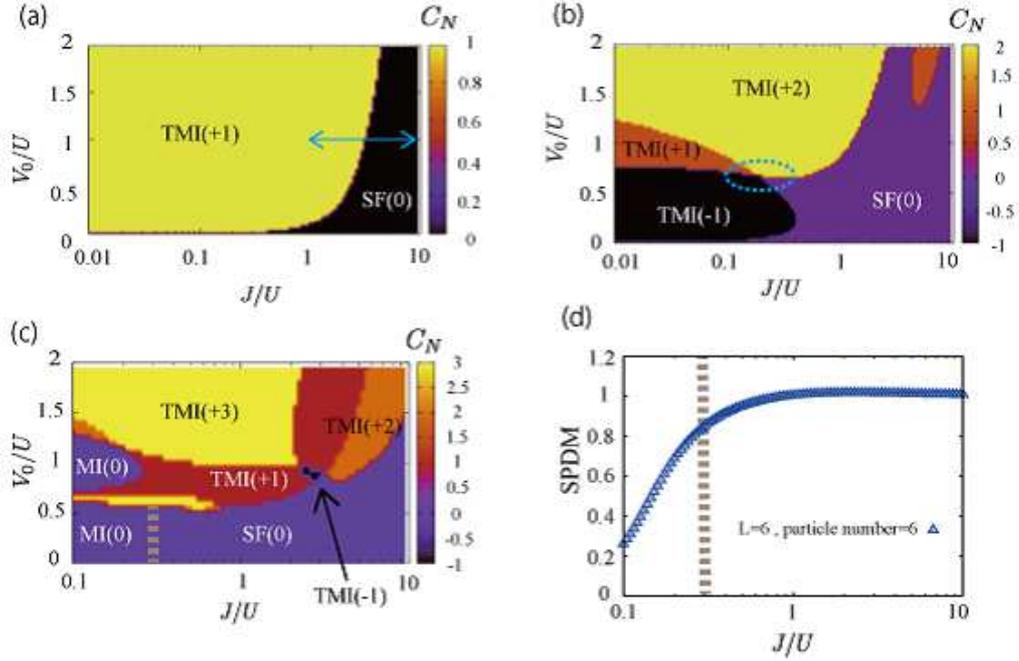}
\caption{Phase diagrams of the soft-core boson system with $\alpha=1/3$. 
(a) $\langle n\rangle = 1/3$ case with $L=12$, 
where the phase diagram is similar to the hard-core case.
Blue arrow in this phase diagram denotes the line on which a finite-size scaling 
analysis is performed (see Fig~.\ref{fs}).
(b) $\langle n\rangle = 2/3$ case with $L=9$, 
the soft-core bosons are allowed to have double occupancy. 
Thus, the non-trivial TMI states form.
(c) $\langle n\rangle = 1$ case with $L=6$.
(d) SPDM for the line $V_{0}/U=0.25$ at $\langle n\rangle =1$ phase diagram in (c)}
\label{Soft_SF-TMI}.
\end{figure}
\begin{figure}[t]
\centering
\includegraphics[width=8cm]{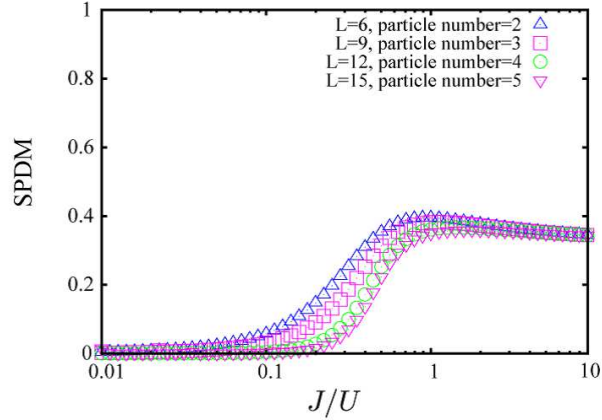}
\caption{The behaviors of SPDM for $V_{0}/U=1$ line in the phase diagram in
Fig.~\ref{Soft_SF-TMI} (a).
Finite $\langle a^{\dagger}_{1}a_{[L/2]}\rangle\neq 0$ indicates that the SF forms there.
We observed that the SPDM has a system-size dependence in most of cases.}
\label{SPDM}
\end{figure}

For the present system, the Chern number is defined as follows on the 2D torus, 
$T^2_{\theta,\delta}$, of the adiabatic parameters $(\theta$, $\delta)$ 
\cite{Zhu,Deng},
\begin{eqnarray}
C_{N}\equiv \frac{i}{2\pi}\int^{2\pi}_{0}\int^{2\pi}_{0} d\theta d\delta \biggl[\partial_{\delta} 
\langle \theta,\delta|\partial_{\theta}|\theta,\delta\rangle- 
\partial_{\theta} \langle \theta,\delta|\partial_{\delta}|\theta,\delta\rangle\biggr],
\label{Cn}
\end{eqnarray}
where 
$|\theta,\delta\rangle$ is the non-degenerate ground state depending on the adiabatic parameters $\theta$ and $\delta$. 

In what follows, we focus on classifying the phase at vanishing temperature
(i.e., the ground state) of the system. 
In the {\it bosonic} system described by the Hamiltonian $H_{\rm BH}$ 
[Eq.(\ref{BHM1D})], three kinds of the ground states \cite{Zhu,Deng,Lang,Xu}
are expected to appear, i.e., superfluid (SF), trivial Mott insulator (MI), and TMI.  
In particular, the TMI is identified as the state with non-vanishing integer $C_{N}$ 
of Eq.(\ref{Cn}) and without SF order. 
The TMI is regarded as an analogous state of the IQH states \cite{Kohmoto,TKNN}. 

We identify the TMIs by using the exact diagonalization \cite{ED1,ED2,ED3}, which 
is an efficient method to study the bulk properties of the system. 
Figure~\ref{Fig2} (b) exhibits our numerical results, i.e., the phase diagram in the
$(\langle n\rangle$ - $V_{0}/J)$-plane for the case of $J=0.01$. 
For calculating the Chern number $C_{N}$ define by Eq.(\ref{Cn}), 
we employed the methods proposed in Ref.~\cite{Fukui} and used the discretized 
adiabatic parameter space with $N\times N$ mesh. 
(We took $N\geq 5$ as suggested in Refs.~\cite{Zhu,Zeng}.) 
As Fig.~\ref{Fig2} (b) shows, the states at specific fillings have a non-vanishing 
Chern number while the others exhibit vanishing Chern number. 
The Chern number $C_{N}$ is quantized as $C_{N}=+1\;(-1)$ at
$\langle n\rangle = 1/3 \; (2/3)$ in the finite-$V_{0}$ region.
Here we note that even for fairly small values of $V_{0}/J$, 
a finite energy gap exists
as seen from Fig.~\ref{Fig2} (a), and this gap accompanies the non-vanishing $C_{N}$. 
For example, the TMI at $\langle n \rangle =2/3$ with $C_N=-1$ forms 
for $V_0/J \ge 0.05$ at which the first-excited energy band appears.
We also exhibit density snapshots for $V_{0}/J=1$ with
$\langle n\rangle=1/3$ and $2/3$ in Figs.~\ref{Fig2} (c) and (d). 
The ground states are non-degenerate and the (discrete) translational invariance is
apparently broken there.
This feature is different from that of the ordinary IQH state, 
in which the  continuous translational invariance is preserved. 

The results obtained in this work are in fairly good agreement with those of 
the previous studies using other numerical methods \cite{Zhu,Deng,Lang,Xu}.  


\section{Phase diagrams of soft-core Bose-Hubbard model}\label{Sec.IV}

In this section, we shall study the soft-core BHM on the superlattice.
In the soft-core case, multiply occupied states are allowed at each site and
therefore the on-site interaction plays an important role to determine 
the ground state of the system. 
In fact, it is expected that the competition between the on-site interaction, $U$, 
and the superlattice, $V_{0}$, leads to rich phase diagrams.
That is, the TMIs with various Chern numbers form in the phase diagram in
the $(J/U-V_0/U)$-plane.
For the numerical methods, we employ the exact diagonalization 
as in the study on the hard-core case. 
In this section, we focus on the global phase diagrams for the cases
$\alpha=1/3,\: 1/4$ and $1/5$.

To begin with, we obtained the phase diagrams in the $(J/U$-$V_{0}/U)$-plane 
with $\alpha=1/3$.
In Figs.~\ref{Soft_SF-TMI} (a) -(c), the phase diagrams of three cases  are shown,
where the particle filling is denoted by $\langle n \rangle$ as before. 

Figure~\ref{Soft_SF-TMI} (a) for the $\langle n \rangle=1/3$ case
shows that there are two phases, i.e., the SF, 
which is characterized by a finite value of the single particle density matrix 
(SPDM) \cite{ED1},  
$\langle a^{\dagger}_{1}a_{[L/2]}\rangle \neq 0$ ($[\ \cdot \ ]$ is the floor function), 
and the TMI phase with $C_{N}=+1$. 
We determined the ground-state phase diagram as follows; 
if the ground state has a non-vanishing $C_{N}$, 
we regard the ground state as a TMI even though the SPDM has a finite value.
That is, we regard the finite SPDM as a finite system-size effect in this case.
(See later discussion.)
On the other hand, if the ground state has a finite SPDM with $C_{N}=0$, 
we regard the ground state as a SF state.
Hereafter, we denote the TMI with the $C_{N}=\pm X$ 
($X=\mbox{positive integer}$) as TMI ($\pm X$).

\begin{figure}[t]
\centering
\includegraphics[width=12cm]{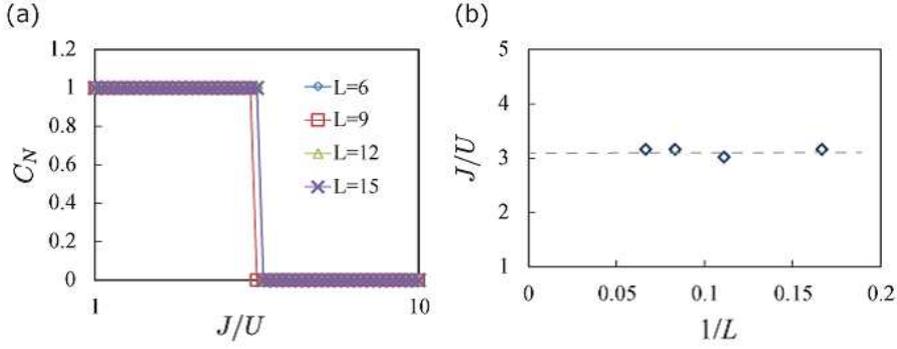}
\caption{ Behavior of the Chern number for $V_{0}/U=1$, 
$\alpha=1/3$ and $\langle n\rangle=1/3$ in Fig.~\ref{Soft_SF-TMI}.
(a) System-size dependence of the Chern number $C_{N}$. 
(b) Finite-size scaling for transition point indicated by the Chern-number $C_{N}$.
The location of the phase transition has a very small system-size dependence.
The  dotted line is for a guide for eyes.}
\label{fs}
\end{figure}

\begin{figure}[t]
\centering
\includegraphics[width=14cm]{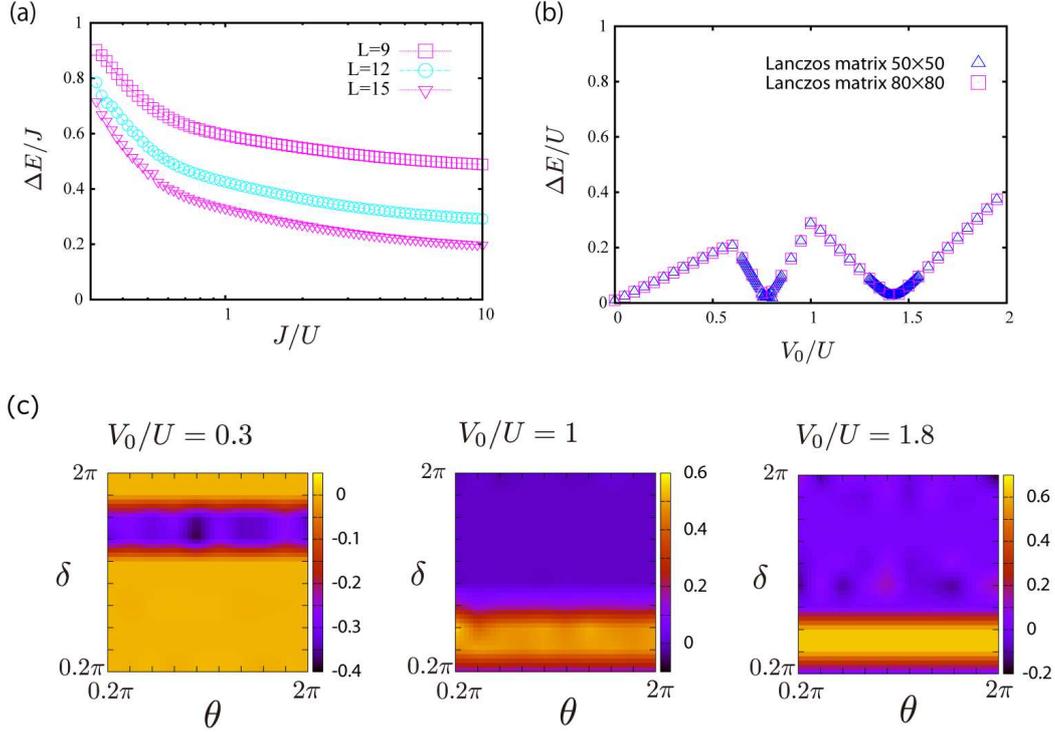}
\caption{(a) Energy gap $\Delta E$ for the TMI(+1)-SF phase transition on the
$V_{0}/U=1$ line.
$\Delta E$ is getting small and the system 
approaches to gapless ground state, i.e., SF as $J/U$ increases. 
However, due to a finite-size effect, the gap does not close even in the SF. 
(b)  $\Delta E$ for the $J/U=0.01$ line in Fig.~\ref{Soft_SF-TMI} (b), 
$\alpha=1/3$, $\langle n\rangle =2/3$. 
The value of $\Delta E$ tends to close on the critical points 
of the topological phase transition. 
(c) Plots of the Berry curvature for the points $V_{0}/U=0.3, \;1.0$ and $1.8$ 
on the parameter space in (b). The mesh size $N=10$.
}
\label{enegap}
\end{figure}

As seen in Fig.~\ref{Soft_SF-TMI} (a), the SF phase forms for large $J/U$. 
The typical behaviors of the SPDM along the $V_{0}/U =1$ line in 
Fig.~\ref{Soft_SF-TMI} (a) are shown in Fig.~\ref{SPDM}. 
The result exhibits the system size-dependence.
On the other hand for the small $J/U$ regime, the system exists in the TMI(+1). 
Figure~\ref{fs} (a) shows the system size dependence of $C_{N}$ 
along the $V_{0}/U=1$ line in Fig.~\ref{Soft_SF-TMI} (a). 
Interestingly, its system size dependence is much smaller than that of the SPDM. 
We also plot the ``finite-size scaling" of the critical point obtained by $C_N$ 
in Fig.~\ref{fs} (b). 
The result indicates that the critical point is almost independent of the system size. 
From the data, we can conclude that the Chern number $C_{N}$ 
is a good order parameter for identifying phase boundaries of the system. 
Similar numerical result concerning to the system size dependence of Chern number 
was reported in Ref.~\cite{Hu}, though the target model is a spin model. 
In addition, we measured the energy gap $\Delta E$ between the ground state and 
the first excited state \cite{Lanczos} along the $V_{0}/U=1$ line in 
Fig.~\ref{Soft_SF-TMI} (a). 
The result is shown in Fig.~\ref{enegap} (a). 
There, gapless excitation in SF cannot be clearly seen
due to the finite-size effect in the small system. 

\begin{figure}[t]
\centering
\includegraphics[width=6cm]{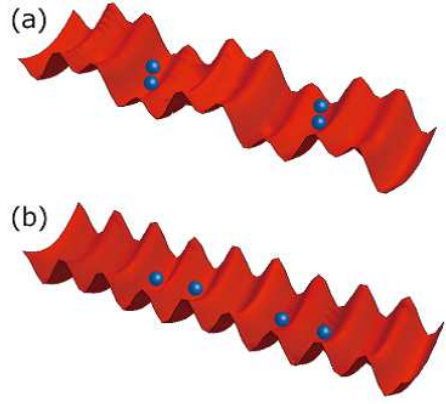}
\caption{(a) Large $V_{0}/U$ case: the system permits double or higher occupancy.
(b) Small $V_{0}/U$ case, bosons tend to exhibit the hard-core nature.
}
\label{double}
\end{figure}

Next, we consider the $\langle n\rangle = 2/3$ case for $\alpha=1/3$.
As shown in Fig.~\ref{Soft_SF-TMI} (b), we obtain the rich phase diagram. 
As far as we know, this phase diagram is one of new findings in this work. 
There are four phases, i.e., the SF phase exists for large $J/U$, 
and interestingly the TMIs with $C_{N}=-1,+1$ and $+2$ appear in the small $J/U$ regime. 
In particular, the TMI with $C_{N}=+2$ is an interesting phase, where  
the system permits the double occupancy at each site as shown in 
Fig.~\ref{double} (a) since $V_{0} >U$, and then the system essentially 
behaves as a two-species boson system.
Each species of boson fully-occupies the lowest band in the SP spectrum in 
Fig.~\ref{Fig2} (a), therefore, $C_{N}=1+1=+2$. 

On the other hand for the case $V_{0} <U$, since the on-site interaction is dominant, 
the multiple occupancy does not appear. 
Bosons behave like fermions as shown in Fig.~\ref{Fig2} (b). 
This situation leads to full-occupancy up to the second-lowest band of 
the SP spectrum. 
Thus, the TMI state exhibits $C_{N}=-1$. 
Similar discussion was given in Ref.~\cite{Deng}. 
Furthermore, we found that the TMI phase with $C_{N}=+1$ forms
as an intermediate state between the two TMIs with $C_{N}=-1$ and $+2$. 
We consider that in this state with $C_N=+1$, the multiply-occupied sites
appear in addition to the fully-occupied lowest band of the SP spectrum. 

Here, we mention the possibility of 
the coexistence of the SF and TMI (non-vanishing $C_{N}$). 
In our simulation, the SPDM certainly exhibits a small but finite value in the TMIs 
near the SF regime.
For example, as one of the possible regime of the coexistence, 
we indicate the area such as 
$0.1\lesssim J/U\lesssim 0.5$ and $0.5\lesssim V_{0}/U\lesssim  0.7$ 
in Fig.~\ref{Soft_SF-TMI} (b) by the dotted ellipse. 
There, a part of bosonic atoms form the TMI(+1) 
and the others may Bose condensate i.e., SF. 
However, due to the finite-size effect, the present calculation of the small system 
cannot reveal the precise behavior of the SPDM.
This problem will be studied in more detail in the near future.

\begin{figure}[t]
\centering
\includegraphics[width=14cm]{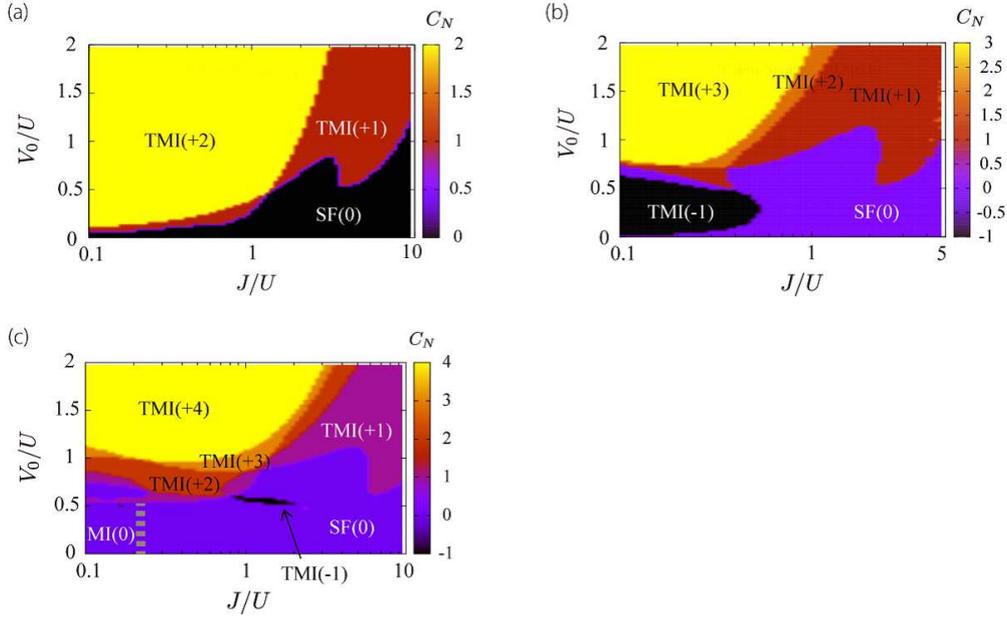}
\caption{Phase diagrams for soft-core case with $\alpha=1/4$. 
(a) $\langle n\rangle=1/2$ and $L=12$.
(b) $\langle n\rangle=3/4$ and $L=8$.
(c) $\langle n\rangle=1$ and $L=4$.
}
\label{Soft_SF-TMIq4}
\end{figure}

\begin{figure}[t]
\centering
\includegraphics[width=14cm]{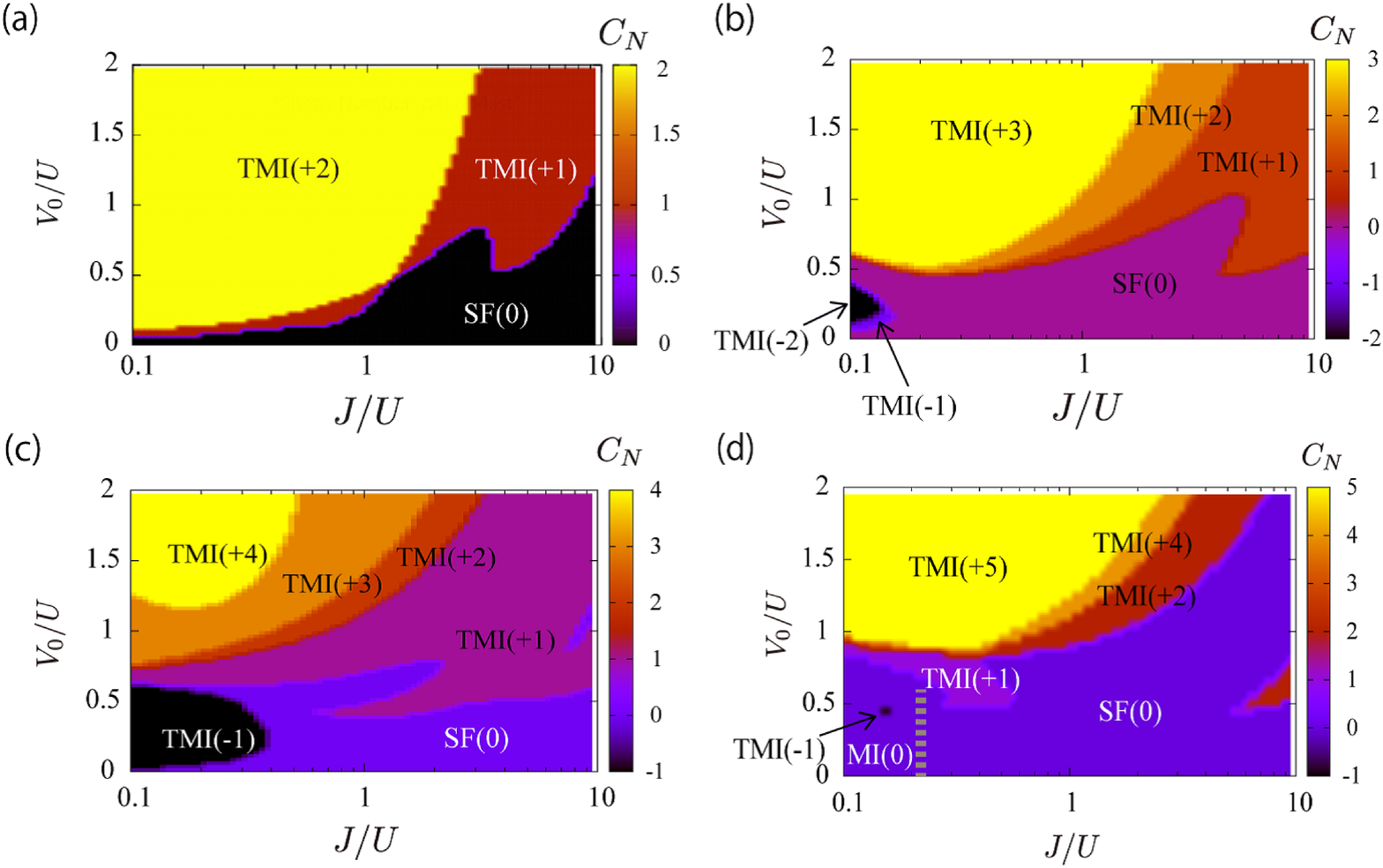}
\caption{
Phase diagrams for soft-core case with $\alpha=1/5$. 
(a) $\langle n\rangle = 2/5$ and $L=10$.
(b) $\langle n\rangle=3/5$ and $L=10$.
(c) $\langle n\rangle=4/5$ and $L=10$.
(d) $\langle n\rangle=1$ and $L=5$.
}
\label{Soft_SF-TMIq5}
\end{figure}

It is interesting to measure the energy gap $\Delta E$ along the lines
in the parameter space, 
on which a phase transition between two different TMIs takes place. 
As a typical example, we plot the energy gap $\Delta E$ 
as a function of $V_{0}/U$ along the $J/U=0.01$ line in the phase diagram 
in Fig.~\ref{Soft_SF-TMI} (b). 
The result is shown in Fig.~\ref{enegap} (b).
For calculation of $\Delta E$, we chose the minimums of $\Delta E$ 
in the adiabatic-parameter space $T^2_{\theta\delta}$.
We find that the gap $\Delta E$ apparently closes at two transition points between 
the TMI(-1) and TMI(+1) and between the TMI(+1) and TMI(+2). 
The result was independent of the size of the tridiagonal matrix of the Lanczos algorithm. 
Our numerical study obviously captures the {\em level-crossing} of the lowest
and first-excited states.
{\em
This level-crossing induces the change of the topological number of the ground state
$C_N$.}

For the $J/U=0.01$ line in the phase diagram 
in Fig.~\ref{Soft_SF-TMI} (b), we show typical distributions of the Berry curvature, 
i.e., the integrand of Eq.(\ref{Cn}).
See Fig.~\ref{enegap} (c).
Its integral over $T^{2}_{\theta\delta}$ gives the non-vanishing integer Chern number.
From the data, the Berry curvature generated 
from the gauge field $i\langle \theta,\delta|\partial_{\mu}|\theta,\delta\rangle$ ($\mu=\theta,\delta$) 
has no topological defect (quantized vortex), but it 
measures the density of the magnetic flux penetrating the surface of the 2D torus 
$T^{2}_{\theta\delta}$.
Then, non-vanishing Chern number indicates the existence of 
a hiding monopole that exists in the {\em interior of the torus} $T^{2}_{\theta\delta}$ 
\cite{Bcarvature}.
In other words, one can define a 3D gauge field induced by $i\langle \theta,\delta|\partial_{\mu}|\theta,\delta\rangle$ 
in the interior of the torus, which corresponds to the magnetic monopole.

Let us see how the phase diagram changes as the filling factor $\langle n\rangle$
increases further.
Figure~\ref{Soft_SF-TMI} (c) shows the phase diagram for the
filling $\langle n\rangle=1$ and $\alpha=1/3$.  
Here, we have a richer phase diagram than the lower filling cases in
Figs.~\ref{Soft_SF-TMI} (a) and (b).
In the phase diagram Fig.~\ref{Soft_SF-TMI} (c), the TMIs with the larger integer
$C_{N}$ form for the large $V_{0}/U$ and small $J/U$ regime. 
In Fig.~\ref{Soft_SF-TMI} (c), the highest value of $C_{N}$ is $+3$. 
This value corresponds to the total particle number in the unit cell, i.e., three particle.
In general, for $\alpha=1/q$ and $\langle n\rangle=K/q$ where $K$ and 
$q$ are co-prime integers, 
the possible highest value of $C_{N}$ is expected to equal the total particle number 
$K$ in the unit cell. 

To get the physical picture of the state with large $C_{N}$, 
we apply the same discussion of the TMI (+2) in Fig.~\ref{Soft_SF-TMI} (b) 
to the present case, 
i.e., the system permits higher occupancy as increasing the value of $V_{0}$ with keeping $U$ fixed.  

Here we remark on the phase boundary of the SF
in the small $V_{0}/U$ and small $J/U$ regime in Fig.~\ref{Soft_SF-TMI} (c).
In that parameter region, we cannot clearly identify 
the phase boundary between the {\em non-topological} MI  and SF state since both
phases have $C_{N}=0$.
Furthermore, the typical behavior of the SPDM shown in Fig.~\ref{Soft_SF-TMI} (d)
indicates that we can obtain the phase boundary only approximately. 
Therefore, we show the approximate phase boundary between the MI and SF 
with the dotted line in Fig.~\ref{Soft_SF-TMI} (c).

We shall show the obtained phase diagrams for other values of 
$\alpha=1/4$ and $1/5$ with various fillings.
Figures~\ref{Soft_SF-TMIq4} (a)-(c) show the global phase diagrams for $\alpha=1/4$.
Similarly to the $\alpha=1/3$ case shown in Fig.~\ref{Soft_SF-TMI}, 
phase diagrams are getting complicated as the filling $\langle n \rangle$ increases.
We note that as increasing $\langle n\rangle$, 
the TMIs with the larger values of  $C_{N}$ appear for large $V_{0}/U$ 
as the on-site energy of the multiply-occupied states is smaller than $V_{0}$ there.
Similarly, we obtained the phase diagrams for the case of $\alpha=1/5$,
and the results are shown in Figs.~\ref{Soft_SF-TMIq5} (a)-(d).
Generally speaking, the phase structures and their dependence on $\langle n\rangle$ 
and $V_{0}$ exhibit similar behaviors with those of the previous cases 
$\alpha=1/3$ and $1/4$.
Unfortunately, the present study using the exact diagonalization 
of the small system is not good enough to get more detailed phase diagrams.
Further numerical studies with large system size, 
e.g., by using the DMRG, is desired.

\begin{figure}[t]
\centering
\includegraphics[width=12cm]{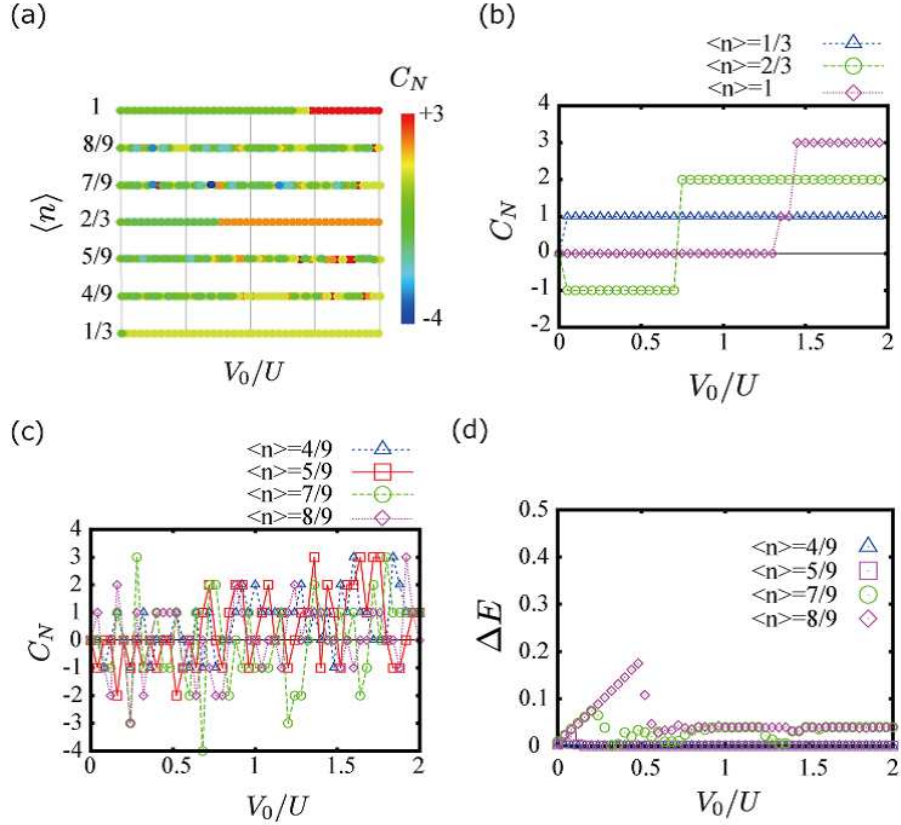}
\caption{
(a) $(\langle n\rangle$-$V_{0}$)-phase diagram of the Chern number $C_{N}$ 
for the soft-core case. $L=9$.
(b) Behavior of the Chern number $C_{N}$ for typical fillings, 
where $C_{N}$ is stable for the change of the parameter $V_{0}/U$.
(c) Behavior of the Chern number $C_{N}$ 
for the cases $\langle n \rangle =4/9, \: 5/9, \: 7/9$ and $8/9$. 
(d) Energy gaps $\Delta E$ for the cases $\langle n \rangle =4/9, \: 5/9, \: 7/9$ and $8/9$.
For $\langle n \rangle =4/9, \: 5/9$, the gaps are very small. 
The level-crossing occurs frequently on varying the parameter $V_{0}/U$. 
Thus, change of the Chern number occurs frequently.
For $\langle n \rangle =7/9,\: 8/9$, the gaps are slightly larger 
than those of the cases $\langle n \rangle =4/9, \: 5/9$.}
\label{Cn_V0}
\end{figure}

Finally, we consider the filling-factor dependence of the system, i.e., the phase diagram in 
the $(\langle n \rangle$-$V_{0}/U)$-plane.
Due to the on-site interaction, the SP picture is no longer meaningful, i.e.,
the energy spectrum of the ground state and the energy gap $\Delta E$ cannot be
described by the SP equation of Eq.~(\ref{SingleE}), even though TMIs may form 
in certain parameter regions.

We obtained the global phase diagram in the $(\langle n\rangle$-$V_{0}/U)$-plane,
which is shown in Fig.~\ref{Cn_V0} (a).
The system size $L=9$, $\alpha=1/3$, and $J/U=0.01$.
The results indicates the existence of two types of non-trivial TMI. 
We call the first one stable TMI (STMI) phase, 
in which $C_{N}$ is robust for the change of the value of $V_{0}/U$ as shown in Fig.~\ref{Cn_V0}~(b).  
On the other hand, we call the second one random Chern number TMI (RTMI) phase, 
in which $C_{N}$ takes various integer values as shown in Fig.~\ref{Cn_V0} (c). 
As seen from Fig.~\ref{Cn_V0} (b), the STMI phase forms
when the filling-factor $\langle n\rangle$ and the modulate parameter $\alpha$ 
are tuned to produce the band-insulator regime, 
and the Chern number is determined 
as in the previous discussion based on the intuitive picture shown in Fig.~\ref{double}. 
On the other hand, the RTMI forms
when the parameters $\langle n\rangle$ and $\alpha$ are not located at 
the band insulator of the SP spectrum, and the interplay of the on-site repulsion
and superlattice plays an essential role there.  
As far as we know, the phase diagram in Fig.~\ref{Cn_V0}~(a) is one of new findings
in this work.

To study detailed properties of the RTMI, we measured the energy gap $\Delta E$ 
for the cases $\langle n\rangle = 4/9,\; 5/9, \: 7/9$ and $8/9$. 
The obtained results are shown in Fig.~\ref{Cn_V0} (d).  
It is obvious that the energy gaps are small for most of the values of $V_{0}/U$. 
This indicates that the RTMI ground states are very close to the first-excited state.
Then, the level-crossing frequently occurs as $V_0/U$ varies and states with 
various Chern number $C_N$ appear as the ground state.


\section{Topological charge pumping and Chern number}\label{Sec.V}

In Sec.~\ref{Sec.IV}, we clarified the ground-state phase diagram of $C_{N}$.
Here we expect that each TMI with different $C_{N}$ exhibits
different transport properties when one varies the cyclic pumping parameter
 $\delta$ adiabatically, e.g., 
as $\delta(t)=2\pi t/T$ ($t\in [0,T]$), 
where $T$ is the period of one pumping cycle.
Such a charge pumping phenomenon was studied 
both theoretically \cite{Thouless,Wei,Wang,Qian,Hatsugai} and experimentally \cite{Nakajima,Lohse,Kraus2} 
for similar models to the present one.
In this section, we shall show a connection 
between the Chern number $C_{N}$ [Eq.(\ref{Cn})] and the particle transfer.

\begin{figure}[t]
\centering
\includegraphics[width=12cm]{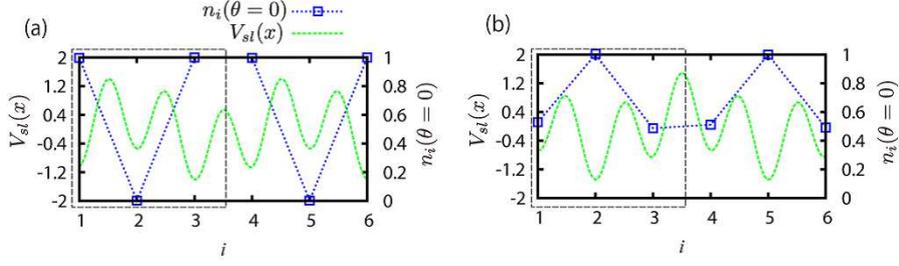}
\caption{Detailed forms of the superlattice potential $V_{sl}(x)$ (the left axis)
and particle density $n_{i}(\theta=0)$ (the right axis).
(a) For $\delta=3\pi/5$, particle is hard to hop to the nearest-neighbor unit cell. 
(b) For $\delta=8\pi/5$, particle is easy to hop to the nearest-neighbor unit cell. 
The boxes denoted by the dashed lines represent one unit cell with $\alpha=1/3$.
The lattice spacing is unity.} 
\label{hopping}
\end{figure}

To begin with, we introduce a current operator,
\begin{eqnarray}
\hat{J}_{c}=\hbar^{-1}\partial_{\theta}H_{\rm BH}
=\frac{1}{\hbar L}\biggl[iJe^{-i\theta/L}\sum_{i}a^{\dagger}_{i}a_{i+1}+\mbox{h.c}\biggr].
\label{Current} 
\end{eqnarray}
From the current operator $\hat{J}_{c}$, 
the current density $\delta I (t)$ at time $t$ can be directly calculated \cite{Hatsugai}.
By using the genuine ground state denoted by $|\Psi(t)\rangle$, 
which satisfies the Schr\"{o}dinger equation
$i\hbar \partial_{t}|\Psi(t)\rangle = H_{\rm BH}|\Psi(t)\rangle$ 
and an adiabatic {\it instantaneous ground state} denoted by $|\psi_{0}(t)\rangle$, 
the current density $\delta I (t)$ is expressed as
\begin{eqnarray}
\delta I(t) = \langle \Psi(t)|\hat{J}_{c}|\Psi(t)\rangle - \langle \psi_{0}(t)|\hat{J}_{c}|\psi_{0}(t)\rangle.
\label{dcurrent} 
\end{eqnarray}
Here, we assume that the many-body interacting system has a finite energy gap and 
introduce energy spectrum $E_{n}(t)$, where the corresponding eigenstate 
is the instantaneous normalized eigenfunctions $|\psi_{n}(t)\rangle$ satisfying
$H_{\rm BH}|\psi_{n}(t)\rangle = E_{n}(t)|\psi_{n}(t)\rangle$. 
For $n=0$, the eigenfunction is nothing but the adiabatic instantaneous ground state. 
Then, the genuine ground state $|\Psi(t)\rangle$ is approximately expanded 
in terms of the states $|\psi_{n}(t)\rangle$ as \cite{Thouless}
\begin{eqnarray}
|\Psi(t)\rangle =&& e^{-\frac{i}{\hbar}\int^{t} E_{0}(t')dt'}\nonumber\\
&&\hspace{-0.2cm}\times
\biggl[ 
|\psi_{0}(t)\rangle +i\hbar \sum_{j\neq 0}
\frac{\langle\psi_{j}(t)|\partial_{t}|\psi_{0}(t)\rangle}{E_{j}(t)-E_{0}(t)}|\psi_{j}(t)\rangle\biggr].
\label{adiabaticexpantion} 
\end{eqnarray}
The derivation of Eq.~(\ref{adiabaticexpantion} ) is given in appendix A.
By substituting the expansion Eq.(\ref{adiabaticexpantion}) into the current density
$\delta I(t)$, Eq.(\ref{dcurrent}), and by using the relation, 
$$
\langle \psi_{0}(t)| \partial_{\theta}H_{\rm BH}|\psi_{j}(t)\rangle = 
-[E_{j}(t)-E_{0}(t)]\langle\partial_{\theta} \psi_{0}(t)| \psi_{j}(t)\rangle,
$$
we obtain the following equation 
(The detailed calculation from Eq. (\ref{dcurrent}) to Eq. (\ref{ddcurrent}) is 
given in appendix B.),
\begin{eqnarray}
\delta I(t) \simeq -i\langle \partial_{\theta}\psi_{0}(t)|\partial_{t}\psi_{0}(t)\rangle 
+ i\langle\partial_{t}\psi_{0}(t)|\partial_{\theta}\psi_{0}(t)\rangle,
\label{ddcurrent} 
\end{eqnarray} 
where we have assumed 
$$|\langle\psi_{j}(t)|\partial_{t}|\psi_{0}(t)\rangle| \ll {E_{j}(t)-E_{0}(t)}.$$
We are interested in the current density $\overline{\delta I(t)}$ averaged over 
$\theta\in [0,2\pi]$ \cite{Niu},
\begin{eqnarray}
\overline{\delta I(t)}=\frac{1}{2\pi}\int^{2\pi}_{0}d\theta \ \delta I(t).
\label{averagecurrent} 
\end{eqnarray} 
Then, we obtain the total particle transfer $\Delta Q$ for one pumping cycle $T$ as
follows,
\begin{eqnarray}
\Delta Q=\int^{T}_{0}dt \ \overline{\delta I(t)}
=\frac{1}{2\pi}\int^{T}_{0}dt \int^{2\pi}_{0} d\theta \ \delta I(t).
\label{totalcurrent} 
\end{eqnarray} 
By introducing the Berry connection $A_\mu$ ($\mu=\theta,t$) by
$i\langle \psi_{0}(t)|\partial_{\mu}|\psi_{0}(t)\rangle= A_{\mu}$, 
$\Delta Q$ is expressed as
\begin{eqnarray}
\Delta Q \simeq\frac{1}{2\pi}\int^{2\pi}_{0}d\theta \int^{T}_{0}dt \ [\partial_{t}A_{\theta}-\partial_{\theta}A_{t}].
\label{totalcurrent2} 
\end{eqnarray}
Finally, by changing variables from $t$ to $\delta$,  $\Delta Q$ is expressed as
\begin{eqnarray}
\Delta Q \simeq \frac{1}{2\pi}\int^{2\pi}_{0}d\theta \int^{2\pi}_{0}d\delta \; 
[\partial_{\delta}A_{\theta}-\partial_{\theta}A_{\delta}], 
\label{DQfinal}
\end{eqnarray}
where $A_{\delta}\equiv i\langle \psi_{0}(t)|\partial_{\delta}|\psi_{0}(t)\rangle$.
From the final expression of $\Delta Q$ in Eq.(\ref{DQfinal}), 
the total particle transfer $\Delta Q$ obviously 
corresponds to the Chern number $C_{N}$ of Eq.(\ref{Cn}), i.e., 
\begin{eqnarray}
\Delta Q \simeq C_{N}.
\label{Q-Cn} 
\end{eqnarray}
The above relation can be regarded as the interacting many-body version 
of the similar relation in the SP picture obtained
by using a single Wannier state and the Bloch wave function \cite{Asboth}, 
which is applicable for a non-interacting SP system.

Here, we note that the data shown in Fig.~\ref{enegap} directly show
the time evolution of $\overline{\delta I(t)}$ because $\delta(t)=2\pi t/T$.
$\overline{\delta I(t)}$ corresponds to the number of the transfer particle to a 
nearest-neighbor unit cell.
The time evolution of $\overline{\delta I(t)}$ originates from the form of 
the superlattice potential 
depending on the parameter $\delta(t)$.  
As one of typical example of the behavior of $\overline{\delta I(t)}$, 
we focus on the TMI (-1) state in the left panel in Fig.~\ref{enegap} (c).
We consider that the superlattice potential can be described 
as $V_{sl}(x)=-V_{s}\cos(2\pi x)+V_{l}\cos(2\pi \alpha x+\delta)$ 
\cite{Nakajima,Lohse}, where the first term is the standard optical lattice potential 
with amplitude $V_{s}$ and the second term is another lattice potential with amplitude $V_{l}$ as shown in Fig.~\ref{Fig1}.
We set a typical ratio $V_{s}/V_{l}=2$ \cite{Lohse}. 
In Fig.\ref{hopping}, we show $V_{sl}(x)$ with $\delta=3\pi/5,\; 8\pi/5$ and 
the corresponding particle density at $\theta=0$ denoted by $n_{i}(\theta=0)$ 
in two nearest-neighbor unit cells. 
From the data, transfer property (hopping tendency) can be intuitively understood. 
In Fig. \ref{hopping} (a), the particle density of the two lattice sites between 
the nearest-neighbor unit cells are almost unity. 
Then, particle is hard to hop between the nearest-neighbor unit cells due to 
strong $U$ and the particle current is suppressed. 
The result directly appears as the small value of the Berry curvature 
at $\delta \sim 3\pi/5$ in the left panel in Fig.~\ref{enegap} (c). 
On the other hand in Fig. \ref{hopping} (b), the particle density of the two lattice 
sites between the nearest-neighbor unit cells are much less than unity and 
particle is easy to hop between the nearest-neighbor unit cells. 
The condition leads to significant particle current. 
The tendency clearly appears in Fig.~\ref{enegap} (c). 
The value of the Berry curvature near $\delta \sim 8\pi /5$ is negatively large, i.e., 
the current to a left nearest-neighbor unit cell is large. 
The total amount of the negative current corresponds to the Chern number ($C_{N}=-1$).

In fact, the current density is determined by the wave functions of the genuine
and instantaneous ground states, but we think that the above observation sheds light 
on the intuitive understanding of the relation between the charge pumping and
the Berry connection for the interacting Bose particle cases.
Here, we should mention the study on the system with edges in Ref.~\cite{Hatsugai}. 
There, focusing on the bulk edge-correspondence in topological charge pumping,
(for related experiments on cold atoms, see Refs.~ \cite{Hadzibabic1,Hadzibabic2})
it was shown that the Berry connection in the temporal gauge is directly  related to the shift
of the center of mass of the system, and the charge pumping is derived by that Berry connection.
For the present system without edges, we think that there exists a direct relation between
the Berry connection and the charge pumping.
This problem is under study, and we hope that the results will be published in the near future.

Equation (\ref{Q-Cn}) shows that TMIs exhibit various amounts of the charge transfer
in the topological charge pumping
depending on the value of the Chern number $C_N$.
Therefore, the obtained phase diagrams of $C_{N}$ in Sec.~IV
can be a guide for detecting new properties of topological charge pumping in 
the strongly-correlated bosonic systems in experiments on ultracold atoms. 


\section{Conclusion}\label{Sec.VI}

In this paper we studied the strongly-interacting boson systems in superlattice potential, 
which are feasible in 1D optical superlattice system of ultracold atoms.
The SP properties of the system are described by the Harper equation, and
the system exhibits the band insulator in the hard-core boson limit.
The band insulator has a non-trivial Chern number, and
it was calculated by obtaining many-body wave functions with varying 
two adiabatic parameters in $T^2_{\theta\delta}$.
For the hard-core boson case with using the exact diagonalization,
we first verified the locations of the band insulator corresponding to the TMI state, 
which is characterized by a non-vanishing Chern number.
Then, we studied the soft-core boson case. 
There, the competition (interplay) between the superlattice amplitude $V_{0}$ and 
on-site interaction $U$ plays an important role.
As a result, we observed the various TMIs with different integer Chern numbers. 
For the modulate parameter $\alpha=1/3,\: 1/4$ and $1/5$, 
we clarified the global phase diagrams of the Chern number 
by using the exact diagonalization. 
The numerical results obtained for different system sizes show that
the behaviors of the Chern number are almost independent of the system size. 
Therefore, the obtained phase diagrams in the present study captures the essential
properties of the system. 
Interestingly, these obtained phase diagrams exhibit a very rich phase structure 
including various TMIs with large Chern numbers.
From the obtained results, we see that the TMI with a high Chern number tends to form 
when the superlattice amplitude $V_{0}$ is getting large.
We conclude that
this behavior originates from the higher occupancy of bosons per lattice site. 
Furthermore,  we clarified particle-filling dependence of the ground state
for $\alpha=1/3$ by studying the $(\langle n\rangle$-$V_{0}/U)$-phase diagram.
We found that there are two types of non-trivial TMI, i.e., the RTMI and the STMI.
The results concerning to the STMI are in fairly good agreement with 
the DMRG result for
a similar model in Ref.~\cite{Deng}, while the existence of the RTMI 
with the interesting behavior of the Chern number is one of new findings of 
the present work.
We also found that level-crossing frequently occurs in the parameter region of the
RTMI.

In Sec.~\ref{Sec.V}, we studied the relationship 
between the particle transfer and the Chern number of many-body wave functions.
We conclude that in order
to detect the various TMIs in real experiments, the charge pumping measurement
by adiabatic time-evolution of the parameter $\delta$ 
\cite{Nakajima,Lohse} is useful.
To measure the TMIs in real experiments, the Streda formula \cite{Streda} is expected
to be useful as pointed out in Ref.~\cite{Zhu}.      

\section*{Acknowledgments}
We acknowledge Y. Takahashi and S. Nakajima for helpful discussions.
Y. K. acknowledges the support of a Grant-in-Aid for JSPS
Fellows (No.17J00486). 

\appendix
\renewcommand{\thefigure}{\Alph{section}.\arabic{figure}}
\setcounter{figure}{0}
\renewcommand{\theequation}{A.\arabic{equation}}

\section*{Appendix A. Derivation of Eq. (\ref{adiabaticexpantion})}

We give the detailed derivation of Eq. (\ref{adiabaticexpantion}) \cite{Asboth}. 
In adiabatic time evolution, the genuine ground ground state $|\Psi(t)\rangle$ 
can be expanded 
by using the instantaneous bases $|\psi_{j}\rangle$, 
which satisfy $H_{BH}(t)|\psi_{j}\rangle=E_{j}(t)|\psi_{j}(t)\rangle$, i.e.,
\begin{eqnarray}
|\Psi(t)\rangle =&& c_{0}(t)e^{-\frac{i}{\hbar}\int^{t} E_{0}(t')dt'}|\psi_{0}(t)\rangle
 +\sum_{j\neq 0}c_{j}(t)e^{-\frac{i}{\hbar}\int^{t} E_{j}(t')dt'}|\psi_{j}(t)\rangle,
\label{adiabaticexpantion0} 
\end{eqnarray}
where $c_{0}(t)$ and $c_{j}(t)$ are time-dependent expansion coefficients.
Since the genuine ground state $|\Psi(t)\rangle$ satisfies
the Shr\"{o}dinger equation $H_{BH}(t)|\Psi(t)\rangle =i\hbar\partial_{t}|\Psi(t)\rangle$ 
and adiabatic setup allows us to approximate 
$c_{0}(t)\simeq 1$, $\partial_{t}c_{0}(t)\simeq 0$ and $|c_{j}(t)|\ll 1$ ($j\neq 0$) along time evolution, we can show that the $j$-th coefficient $c_{j}(t)$ satisfies  
the following time differential equation, 
\begin{eqnarray}
\partial_{t}c_{j}(t)\simeq -e^{\frac{i}{\hbar}\int^{t} (E_{j}(t')-E_{0}(t'))dt'}\langle\psi_{j}(t)|\partial_{t}|\psi_{0}(t)\rangle.
\label{coefficientj} 
\end{eqnarray}  
Then, the adiabatic solution to Eq.~(\ref{coefficientj}) in 
$\mathcal{O}(|\langle\psi_{j}(t)|\partial_{t}|\psi_{0}(t)\rangle/[E_{j}(t)-E_{0}(t)]|)$ 
is obtained as \cite{Asboth},  
\begin{eqnarray}
c_{j}(t)\simeq i \hbar e^{\frac{i}{\hbar}\int^{t} (E_{j}(t')-E_{0}(t'))dt'}\frac{\langle\psi_{j}(t)|\partial_{t}|\psi_{0}(t)\rangle}{E_{j}(t)-E_{0}(t)}.
\label{coefficientjj} 
\end{eqnarray}
By substituting the coefficients in Eq.~(\ref{coefficientjj}) into 
Eq.~(\ref{adiabaticexpantion0}) we obtain the adiabatic expansion
Eq.~(\ref{adiabaticexpantion}) in Sec.~5.  

\section*{Appendix B. Derivation of Eq. (\ref{ddcurrent})}

We give the detailed calculation from Eq. (\ref{dcurrent}) to Eq. (\ref{ddcurrent}).
By substituting Eq. (\ref{adiabaticexpantion}) into Eq.(\ref{dcurrent}), the current $\delta I$ is expressed as, 
\begin{eqnarray}
\delta I(t) &=& \langle \Psi(t)|\hat{J}_{c}|\Psi(t)\rangle - \langle \psi_{0}(t)|\hat{J}_{c}|\psi_{0}(t)\rangle,\nonumber\\
&=& i\sum_{j\neq 0} \frac{\langle \psi_{j}(t)|\partial_{t}|\psi_{0}(t)\rangle \langle\psi_{0}(t)|\partial_{\theta}H_{\rm BH}|\psi_{j}(t)\rangle}{E_{j}(t)-E_{0}(t)}\nonumber\\
&&-i\sum_{j\neq 0} \frac{\langle \partial_{t} \psi_{0}(t)|\psi_{j}(t)\rangle \langle\psi_{j}(t)|\partial_{\theta}H_{\rm BH}|\psi_{0}(t)\rangle}{E_{j}(t)-E_{0}(t)} \nonumber\\
&&+\mathcal{O}\biggl(\biggl|\frac{\langle\psi_{j}(t)|\partial_{t}|\psi_{0}(t)
\rangle}{E_{j}(t)-E_{0}(t)}\biggr|^{2}\biggr),
\label{expantion1} 
\end{eqnarray}
where the term $\langle \psi_{0}(t)|\hat{J}_{c}|\psi_{0}(t)\rangle$ was canceled.
Here we drop the last term in Eq.~(\ref{expantion1}) since we assume the energy gap 
between the ground state and the excited states, $E_{j}(t)-E_{0}(t)$, is large enough. 
Then we notice the following equations, 
$$
\langle \psi_{0}(t)| \partial_{\theta}H_{\rm BH}|\psi_{j}(t)\rangle = 
-[E_{j}(t)-E_{0}(t)]\langle\partial_{\theta} \psi_{0}(t)| \psi_{j}(t)\rangle,
$$
$$
\langle \psi_{j}(t)| \partial_{\theta}H_{\rm BH}|\psi_{0}(t)\rangle = 
[E_{j}(t)-E_{0}(t)]\langle \psi_{j}(t)|\partial_{\theta} \psi_{0}(t)\rangle.
$$
The above equations are obtained from 
$\partial_{\theta}\langle \psi_{0}(t)|\psi_{j}(t)\rangle=0$ and 
$\partial_{\theta}\langle \psi_{j}(t)|\psi_{0}(t)\rangle=0$. 
Then substituting the above equations in Eq. (\ref{expantion1}) 
and using the complete relation of the state bases 
$\sum_{j}|\psi_{j}(t)\rangle\langle \psi_{j}(t)|=1$, 
the current $\delta I(t)$ is expressed as 
\begin{eqnarray}
\delta I(t) \simeq -i\langle \partial_{\theta}\psi_{0}(t)|\partial_{t}\psi_{0}(t)\rangle 
+ i\langle\partial_{t}\psi_{0}(t)|\partial_{\theta}\psi_{0}(t)\rangle.
\label{dddcurrent} 
\end{eqnarray}



\begin{thebibliography}{10}
\bibitem{Jordan}
Rigorously, the connection is achieved though the Jordan-Wigner transformation.
\bibitem{Lanczos}
In our exact diagonalization, the energy gap $\Delta E$ can be obtained from a tridiagonal matrix 
with the optimal size of the dimension from the Lanczos algorithm.
\bibitem{Bcarvature}
A similar results are plotted in Ref.\cite{Fukui}.
\end{thebibliography}
\end{document}